\documentclass[sigconf,screen, pbalance]{acmart}
\AtBeginDocument{%
  }

\usepackage{enumitem}
\usepackage{etc}
\usepackage{listings}
\usepackage{xcolor}

\definecolor{solarBase}{RGB}{238, 243, 238}
\definecolor{solarKey}{RGB}{181, 137, 0}   
\definecolor{solarString}{RGB}{44, 44, 44} 
\definecolor{solarNum}{RGB}{220, 50, 47}     

\lstdefinelanguage{jsonstyle}{
    basicstyle=\scriptsize\ttfamily,
    backgroundcolor=\color{solarBase},
    stringstyle=\color{solarString},
    numbers=none, 
    frame=single,
    framerule=0pt, 
    framesep=3pt,       
    aboveskip=5pt,      
    belowskip=5pt,      
    columns=flexible,   
    breaklines=true,
    showstringspaces=false,
    string=[s]{"}{"},
    literate=
     *{0}{{{\color{solarNum}0}}}{1}
      {1}{{{\color{solarNum}1}}}{1}
      {2}{{{\color{solarNum}2}}}{1}
      {3}{{{\color{solarNum}3}}}{1}
      {4}{{{\color{solarNum}4}}}{1}
      {5}{{{\color{solarNum}5}}}{1}
      {6}{{{\color{solarNum}6}}}{1}
      {7}{{{\color{solarNum}7}}}{1}
      {8}{{{\color{solarNum}8}}}{1}
      {9}{{{\color{solarNum}9}}}{1}
}

\lstdefinelanguage{bashstyle}{
    basicstyle=\small\ttfamily,
    backgroundcolor=\color{solarBase},
    stringstyle=\color{solarString},
    numbers=none, 
    frame=single,
    framerule=0.5pt, 
    framesep=3pt,       
    columns=flexible,   
    breaklines=true,
    showstringspaces=false,
    string=[s]{"}{"},
    literate=
     *{0}{{{\color{solarNum}0}}}{1}
      {1}{{{\color{solarNum}1}}}{1}
      {2}{{{\color{solarNum}2}}}{1}
      {3}{{{\color{solarNum}3}}}{1}
      {4}{{{\color{solarNum}4}}}{1}
      {5}{{{\color{solarNum}5}}}{1}
      {6}{{{\color{solarNum}6}}}{1}
      {7}{{{\color{solarNum}7}}}{1}
      {8}{{{\color{solarNum}8}}}{1}
      {9}{{{\color{solarNum}9}}}{1}
}

\setlist{topsep=0pt, leftmargin=*}
\setcopyright{none}
\renewcommand\footnotetextcopyrightpermission[1]{} 

\begin{document}

\title{CodeGreen: Towards Improving Precision and Portability in Software Energy Measurement}

\author{Saurabhsingh Rajput}
\affiliation{
  \institution{Dalhousie University}
  \city{Halifax}
  \country{Canada}}
\email{saurabh@dal.ca}

\author{Tushar Sharma}
\affiliation{
  \institution{Dalhousie University}
  \city{Halifax}
  \country{Canada}}
\email{tushar@dal.ca}

\renewcommand{\shortauthors}{Rajput et al.}


\begin{abstract}
Accurate software energy measurement is critical for optimizing energy, yet existing profilers force a trade-off between measurement accuracy and overhead due to tight coupling with supported specific hardware or languages.
We present \textbf{CodeGreen}, a modular energy measurement platform that decouples instrumentation from measurement via an asynchronous producer-consumer architecture. We implement a Native Energy Measurement Backend (NEMB) that polls hardware sensors (Intel RAPL, NVIDIA NVML, AMD ROCm) independently, while lightweight timestamp markers enable tunable granularity.
CodeGreen leverages Tree-sitter AST queries for automated instrumentation across Python, C++, C, and Java, with straightforward extension to any Tree-sitter-supported grammar, enabling developers to target specific scopes (loops, methods, classes) without manual intervention. Validation against \emph{Computer Language Benchmarks Game} demonstrates $R^2 = 0.9934$ correlation with RAPL ground truth and $R^2 = 0.9997$ energy-workload linearity. By bridging fine-grained measurement precision with cross-platform portability, CodeGreen enables practical algorithmic energy optimization across heterogeneous environments. Source code~\cite{Rajput_CodeGreen_A_Modular_2026}, video demonstration~\cite{codegreen_demo}, and documentation~\cite{codegreen_website} for the tool are publicly available.
\end{abstract}


\maketitle

\section{Introduction}

The cessation of Dennard scaling~\cite{Dennard1974} and the surge in energy-intensive AI workloads~\cite{epoch2025howmuchenergydoeschatgptuse} have transformed software energy consumption into a critical constraint on operational costs and environmental sustainability. 
While the ``Green Software''~\cite{verdecchia2021green} movement posits that code itself---its logic, algorithms, and architectural patterns---is a primary driver of carbon emissions, the practical optimization of these software artifacts faces a critical tooling gap:
\textit{measuring energy consumption at source-code granularity without compromising measurement validity}. Over the past decade, substantial progress has been made in developing energy measurement tools, yet the majority remain tailored for research environments, requiring specialized hardware access, manual instrumentation, and domain-specific expertise. These barriers impede widespread adoption among software developers, who are the primary agents of energy optimization at the algorithmic and architectural levels.

Existing tools often presume homogeneous execution environments, targeting specific hardware (\eg{} Intel RAPL~\cite{khan2018rapl}) or specific languages (\eg{} Python frameworks). However, modern software development is characterized by extreme heterogeneity: a codebase may execute across diverse hardware platforms (Intel, AMD, NVIDIA GPUs, ARM processors) while hosting applications written in multiple languages (Python, C++, Java). A developer-centric energy profiling tool must therefore be agnostic to hardware and programming language. Simultaneously, developers seeking to identify energy hotspots during iterative development require fine-grained visibility into individual APIs and methods. This necessity amplifies a fundamental tension in software metrology: the ``\textit{Granularity-Overhead}'' trade-off \ie{} as measurement targets smaller execution units (individual functions or loops), the overhead of the measurement instrument grows disproportionately, rendering data statistically invalid. Therefore, an energy measurement tool must deliver high measurement accuracy while minimizing noise and overhead.

Current tools exhibit critical limitations. HPC-grade profilers (PAPI~\cite{weaver2012measuring}, Likwid~\cite{treibig2010likwid}, EnergAt~\cite{energat}) achieve microsecond resolution but require root access and manual instrumentation. ML-focused tools (CodeCarbon~\cite{codecarbon}, Eco2AI~\cite{budennyy2022eco2ai}) prioritize ease of use but sacrifice accuracy through coarse 15-second sampling and TDP heuristics. Cloud-native tools such as Kepler~\etal{}~\cite{amaral2023kepler} have negligible overhead but operate at process granularity, lacking function-level visibility. Language-specific profilers (PyRapl~\cite{fieni2024}) provide fine-grained attribution but incur prohibitive overhead from synchronous hardware reads and runtime contention. Hardware support remains fragmented: fine-grained profilers couple tightly to vendor interfaces (Intel RAPL), while GPU tools lack unified CPU correlation, forcing developers to manage disparate toolchains. 
These tools reveal a consistent gap: developer-focused profiling requires high temporal resolution and control over granularity without measurement artifacts, automated instrumentation, low and predictable overhead, and multi-hardware and multi-language capability.

To address these limitations, we present \textbf{CodeGreen}, a modular energy measurement platform architected for hardware and language extensibility. CodeGreen separates instrumentation from measurement: applications record lightweight timestamp markers via Tree-sitter-based~\cite{max_brunsfeld_2025_17921700} AST analysis, enabling seamless language extension. The \textbf{Native Energy Measurement Backend (NEMB)} asynchronously polls diverse hardware sensors (Intel RAPL, NVIDIA NVML, AMD ROCm) through an extensible driver interface, enabling precise multi-language profiling across heterogeneous environments. Validation demonstrates $R^2 = 0.9934$ correlation with RAPL ground truth and $R^2 = 0.9997$ energy-workload linearity. CodeGreen is available as an open-source toolkit~\cite{Rajput_CodeGreen_A_Modular_2026} with documentation~\cite{codegreen_website}, targeting developers optimizing algorithmic energy consumption across diverse deployment environments.


\section{Background and Related Work}

\noindent \textbf{The Physics of Software Energy.}
Software energy profiling maps logical program execution to physical power consumption. Unlike intrinsic metrics such as execution time, energy ($E$) is a physical quantity derived from the integral of instantaneous power ($P$) consumed over time: $E = \int_{t_0}^{t_1} P(t) \, dt$. Digital power consumption is a continuous signal driven by transistor switching. To reliably reconstruct this signal, the sampling frequency must satisfy the \textit{Nyquist-Shannon theorem}~\cite{Shannon1949}, mandating that the sampling rate $f_s > 2f_{\text{max}}$, where $f_{\text{max}}$ represents the highest frequency component of the power waveform. However, hardware sensors impose physical ceilings; RAPL updates at $\sim$1 kHz (1 ms intervals). Profilers that sample significantly slower than this hardware limit (\eg{} CodeCarbon's default 15s) are mathematically guaranteed to \textit{alias}~\cite{aliasing2025} the signal, missing transient power spikes caused by bursty workloads.

\noindent \textbf{The Observer Effect.}
Measurement introduces computational perturbations to the system under observation. Synchronous profilers (\eg{} pyRAPL, pyJoules) halt application threads to read hardware registers via kernel context switches. This flushes instruction pipelines, evicts cache data, and prevents low-power CPU states. For fine-grained measurements, this overhead can dominate execution time, invalidating energy attribution.

\noindent \textbf{Instrumentation Strategies.}
\textit{Manual instrumentation} (\eg{} PMT~\cite{corda2022pmt}, Likwid~\cite{treibig2010likwid}) requires developers to explicitly annotate code regions, imposing high cognitive burden and placement error risk. \textit{Automated instrumentation} uses compiler or AST analysis to inject checkpoints programmatically~\cite{rajput2024enhancing}. However, naive automation can lead to over-instrumentation, resulting in unexpected measurement overheads that distort workload behavior. Effective instrumentation balances coverage against intrusion.

\noindent \textbf{Hardware Interface Landscape.}
\emph{CPU Telemetry.} Intel RAPL exposes cumulative energy counters for Package, Cores, and DRAM domains. While RAPL provides reasonable resolution,it suffers from counter wraparound under high load and security restrictions (\eg{} post-Platypus attacks), forcing tools to use slower filesystem interfaces. AMD and ARM offer similar capabilities via \texttt{amd\_energy} and \texttt{hwmon}, but with varying granularity and update frequencies.
\emph{GPU Telemetry.} NVIDIA NVML provides GPU power monitoring but exhibits significant limitations. Yang et al.~\cite{Yang2024} demonstrate that slow averaging windows (1s on consumer GPUs) and default polling rates miss up to 75\% of kernel execution behavior. PowerSensor3~\cite{vandervlugt2025powersensor3fastaccurateopen} confirms NVML consistently underestimates energy.


\subsection{Comparative Tool Analysis}

Existing tools present distinct trade-offs between precision, overhead, and usability.

\noindent \textbf{HPC-Grade Profiling.} PAPI~\cite{weaver2012measuring} and Likwid~\cite{treibig2010likwid} achieve low-latency profiling via direct hardware access but require root privileges and manual annotation. PMT~\cite{corda2022pmt} abstracts heterogeneous accelerators but incurs instrumentation overhead due to a lack of automation support across languages and granularity.

\noindent \textbf{Python/ML Ecosystem.} CodeCarbon~\cite{codecarbon} prioritizes ease of use with coarse 15s sampling intervals. Fischer~\etal{}~\cite{fischer2025groundtruthing} demonstrate this underestimates energy by 20--40\% for bursty workloads due to peripheral power exclusion and TDP heuristics. pyJoules~\cite{Belgaid_Pyjoules_Python_library_2019}, and pyRAPL~\cite{fieni2024}  offer decorator-based profiling but suffers from synchronous I/O and GIL contention. Rajput~\etal{}~\cite{rajput2024enhancing} provide energy measurement limited to Deep Learning Framework APIs.

\noindent \textbf{Language-Specific Agents.} JoularJX~\cite{noureddine2022} provides Java-specific profiling via agent-based stack sampling but introduces up to 8\% overhead and $\pm$5\% attribution error~\cite{joularjx_blog}, as it infers energy from CPU demand rather than direct measurement. 

\noindent \textbf{Model-based Attribution.} SmartWatts~\cite{fieni2020smartwatts} correlates hardware performance counters with energy consumption but requires calibration times of minutes to days per platform~\cite{fieni2024} and mandates a complex stack (MongoDB, InfluxDB, sensor daemon), which is impractical for fine-granular profiling.

\noindent \textbf{Infrastructure Middleware.} Kepler~\cite{amaral2023kepler} and Scaphandre~\cite{scaphandre} use ratio models---energy scales linearly with CPU utilization---
suitable for billing and monitoring, but lacks method-level resolution.


\section{Design and Implementation}

CodeGreen differentiates itself through a design philosophy centered on hardware and language extensibility while minimizing measurement noise through asynchronous decoupling and predictable overhead modeling.

\begin{figure}[ht]
    \centering
    \includegraphics[width=\columnwidth]{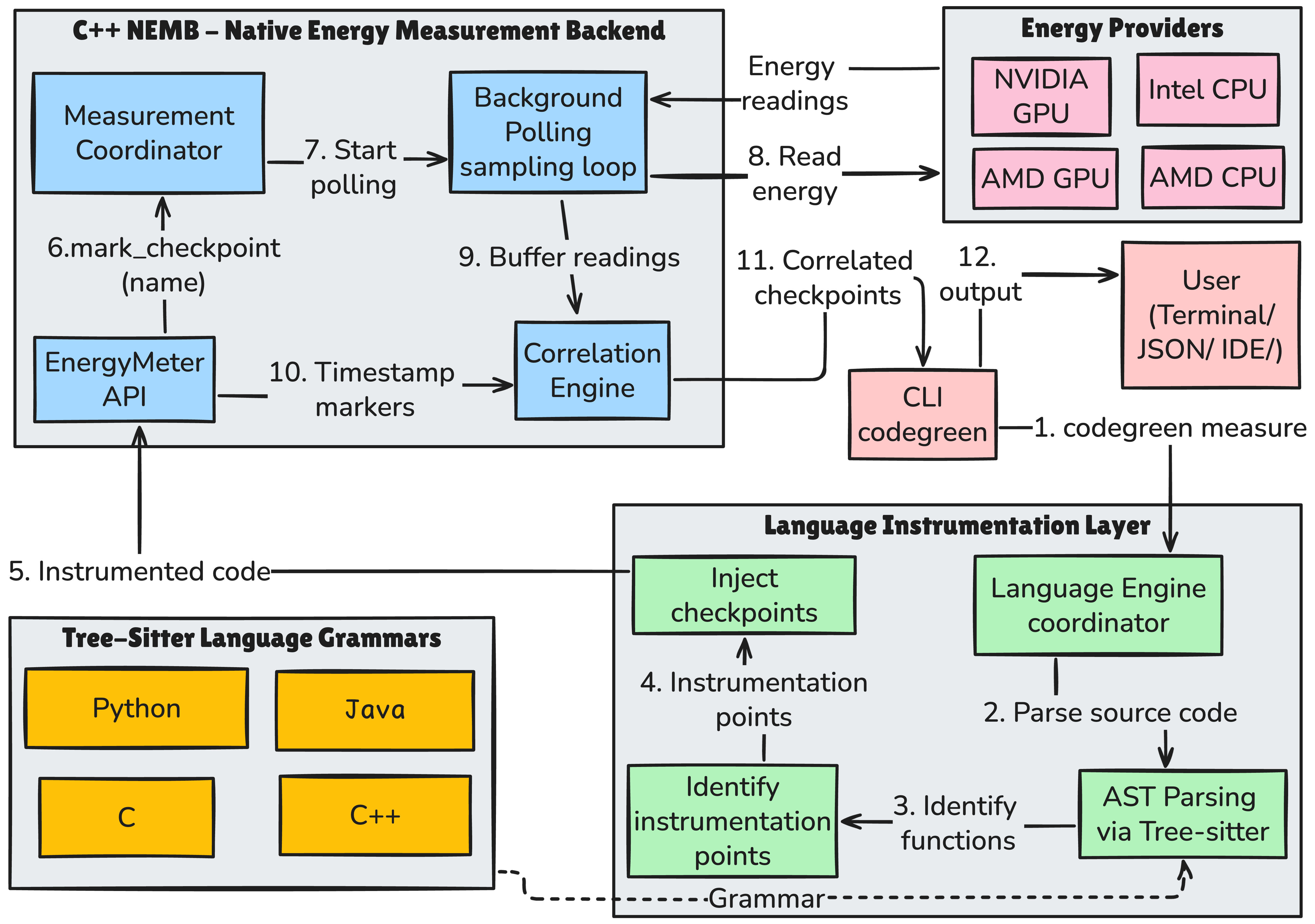}
    \caption{CodeGreen architecture
    }
    \vspace{-3mm}
    \label{fig:arch}
\end{figure}

\noindent \emph{\textbf{Asynchronous Producer-Consumer Architecture.}}
 The core architectural principle of CodeGreen is the \textbf{asynchronous decoupling} of the application under measurement (Producer) from the energy measurement subsystem (Consumer), illustrated in Figure~\ref{fig:arch}. Drawing on event-driven architectures~\cite{eugster2003many}, CodeGreen eliminates the temporal dependency between execution and measurement. Traditional profilers entangle these concerns by blocking the application thread while synchronously querying hardware sensors. In contrast, CodeGreen's application thread performs only a lightweight operation: writing a timestamp marker to a lock-free, thread-safe circular buffer, enabling thousands of checkpoints per second without perturbing the application's energy profile.

The measurement workflow proceeds in three phases, as steps shown in Fig.~\ref{fig:arch}: (1)~\textbf{Instrumentation:} The Language Engine parses source code using Tree-sitter, identifies instrumentation points via SCM queries (steps 2-3), and injects checkpoint calls (step 4) that invoke the \texttt{EnergyMeter} API---the runtime interface responsible for registering execution boundaries. During execution, each checkpoint records only a high-resolution timestamp and unique identifier (step 6), incurring minimal overhead. (2)~\textbf{Background Polling:} A dedicated C++ thread continuously polls hardware sensors at \textit{configurable intervals} (default 10ms, step 7-9), storing \texttt{(timestamp, cumulative\_energy)} tuples in a lock-free buffer. This thread operates independently of the application, preventing measurement operations from blocking execution. (3)~\textbf{Correlation:} Post-execution, the Correlation Engine (step 10-11) uses binary search and linear interpolation to map asynchronous checkpoints onto the energy time series, attributing consumption to code regions.

\noindent \emph{\textbf{Hardware and Language Extensibility}}
CodeGreen implements a modular driver architecture in the C++ backend (Figure~\ref{fig:arch}, Energy Providers). The system defines a generic \texttt{EnergyProvider} interface abstracting vendor-specific hardware APIs. Integrating new platforms (\eg{} AMD ROCm, ARM energy counters) requires only implementing a driver class conforming to this interface without modifying core measurement logic. Current drivers include: \textbf{Intel/AMD RAPL} accessing CPU energy counters via Model Specific Registers (MSR) or Linux \texttt{perf}, supporting Package, Core (PP0), and DRAM domains; \textbf{NVML} monitoring GPU power via the NVIDIA Management Library, reporting Total Graphics Power with domain-level breakdown. The Measurement Coordinator(Figure~\ref{fig:arch}) orchestrates multi-provider sampling, enabling simultaneous CPU and GPU profiling with synchronized timestamps.

Language support leverages the Language Instrumentation Layer (bottom of Figure~\ref{fig:arch}), which uses Tree-sitter parsers to analyze source code via pattern-matching queries (SCM queries). The Language Engine coordinates instrumentation across Python, C, C++, and Java by invoking language-specific Tree-sitter grammars, identifying instrumentation points (\eg{} functions, loops, classes), and injecting checkpoint calls. Extending to additional languages requires only providing the corresponding Tree-sitter grammar. The measurement backend exposes a unified \texttt{EnergyMeter} API that language bindings invoke to mark checkpoints, ensuring consistent behavior across runtimes. Critically, instrumentation granularity is fully configurable: users define targeted scopes---from API endpoints to specific internal loops---via configuration files, precisely calibrating profiling depth to optimization targets.

\noindent \emph{\textbf{Checkpointing and Thread Safety}}
To support complex execution patterns including recursion and multithreading, each checkpoint uses a composite key: \texttt{function\_name\#inv\_N\_tTHREADID}, where \texttt{inv\_N} distinguishes recursive invocations and \texttt{tTHREADID} isolates concurrent threads. This lightweight scheme enables dense instrumentation at function boundaries, loop iterations, and critical sections while maintaining thread safety without locks.

\noindent \emph{\textbf{Predictable Overhead and Noise Reduction}} \label{par:noise}
Measurement accuracy depends on both sensor precision and attribution correctness. CodeGreen enforces temporal consistency by timestamping all events---both application checkpoints and hardware sensor readings---using \texttt{CLOCK\_MONOTONIC}~\cite{clock_monotonic_linux}, a monotonically increasing system clock immune to administrative adjustments or NTP synchronization~\cite{ntp2026}. This unified time reference enables the Correlation Engine to precisely align execution markers with energy measurements, attributing energy to specific code blocks even when hardware sensor update intervals (typically 1ms for RAPL) exceed individual function execution times.

Critically, the continuous polling strategy ensures predictable, constant overhead independent of application behavior. This deterministic behavior enables overhead normalization: since both the fixed initialization cost ($T_{\text{base}}$) and per-checkpoint cost ($T_{\text{checkpoint}}$) are constant for a given runtime, instrumentation overhead can be precisely subtracted from measurements using the model: $\text{Overhead}_{\text{norm}} = (T_{\text{inst}} - T_{\text{native}} - T_{\text{base}} - N \times T_{\text{checkpoint}})/T_{\text{native}} \times 100\%$, where $N$ is the checkpoint count. This normalization yields true measurement overhead after accounting for instrumentation artifacts. Users can thus predict and remove instrumentation costs when interpreting energy measurements, obtaining energy consumption that closely approximates uninstrumented execution.

\noindent \emph{\textbf{Data Correlation and Output}}
Hardware sensor updates occur at fixed intervals (typically 1ms for RAPL) that rarely align with checkpoint timestamps. The Correlation Engine addresses this through temporal interpolation: given energy readings $(t_i, E_i)$ and $(t_{i+1}, E_{i+1})$ and a checkpoint at time $t_c$ where $t_i < t_c < t_{i+1}$, energy is estimated as $E(t_c) = E_i + (E_{i+1} - E_i) \cdot \frac{t_c - t_i}{t_{i+1} - t_i}$. This linear interpolation accurately captures cumulative energy for steady-state workloads. Users configure measurement granularity via CLI (\texttt{codegreen measure --interval 1ms}), trading temporal resolution for storage overhead. Output is structured as JSON (Listing~\ref{lst:output_json}), providing per-checkpoint energy attribution, aggregate statistics, and raw time-series data for external analysis.


\begin{lstlisting}[language=jsonstyle, caption={CodeGreen hierarchical energy attribution output.}, label={lst:output_json}]
{ 
    "timestamp": "2026-01-16T14:30:00.123456",
        "analysis": {
            "language": "python",
            "success": true,
            "instrumentation_points_count": 5,
        },
        "measurement": {
            "total_energy_joules": 1.234,
            "total_time_seconds": 0.567,
            "checkpoints": [
                {   "checkpoint_id": "my_func#inv_1_t123",
                    "type": "function_entry",
                    "energy_joules": 0.001,
                    "time_seconds": 0.0001 },
                {   "checkpoint_id": "loop_body#inv_1_t123",
                    "type": "loop_iteration",
                    "energy_joules": 0.0002,
                    "time_seconds": 0.00005 }, // ... more checkpoints ] 
            }
}
\end{lstlisting}


\section{Validation and Benchmarking}

We validate CodeGreen's measurement accuracy and configurable overhead using the Computer Language Benchmarks Game (CLBG)~\cite{benchmarks_game}, which provides standardized CPU-intensive workloads across Python, C, C++, and Java. We deliberately selected short-running benchmarks because they represent the most challenging measurement scenario: smaller scripts with finer granularity amplify instrumentation overhead relative to execution time, requiring the tool to demonstrate accurate attribution even under adverse conditions.

\noindent \emph{\textbf{Experimental Setup}} 
Experiments were run on an AMD Zen 4 processor, monitoring RAPL package energy. Protocol involved following the best practices from literature~\cite{rajput2024enhancing}, 3 warmup iterations and 30 measured iterations per configuration, reporting mean and 95\% confidence intervals. We evaluated three benchmarks across four languages: \textbf{nbody} (compute-heavy), \textbf{fannkuchredux} (CPU-bound), and \textbf{binarytrees} (memory-intensive), testing CodeGreen at both coarse (project-level) and fine (method-level) granularities.

\noindent \emph{\textbf{Instrumentation Overhead and Checkpoint Granularity}}
For coarse-grained instrumentation with only 2 checkpoints (program entry and exit), raw overhead appears high for fast-running benchmarks: C/C++ show 311--322\% overhead for 40ms execution. However, as discussed in Section~\ref{par:noise}, this is dominated by the fixed 125ms library loading cost, which becomes negligible as program execution time increases. After applying overhead normalization (Sec.~\ref{par:noise}), the true measurement overhead is only 9.1--9.6\% for C/C++, while Python and Java show negligible overhead as their longer runtimes (2.92s and 103ms respectively) amortize the fixed cost.

For fine-grained method-level instrumentation, overhead scales linearly with checkpoint count: \textbf{fannkuchredux} with 4 checkpoints incurs 1.5--11.3\% normalized overhead, \textbf{binarytrees} with 20 checkpoints shows 7.3--22\%, and \textbf{nbody} with 1 million checkpoints exhibits 607\% overhead (C)---an intentional stress test demonstrating per-invocation measurement capability. This deterministic scaling enables users to tune granularity based on optimization needs: coarse-grained measurement (2 checkpoints) provides low-overhead whole-program profiling comparable to \texttt{perf}, while fine-grained instrumentation offers method-level attribution unavailable in process-level tools, accepting higher overhead for detailed hotspot identification. 
Unlike decorator-based profilers, whose overhead is unpredictable and compounds with execution complexity, CodeGreen's predictable, near-constant cost enables precise prediction and control of measurement impact.

\noindent \emph{\textbf{Measurement Accuracy}}
We compare CodeGreen's energy measurements against Linux \texttt{perf}, which serves as ground truth by reading directly from RAPL Model Specific Registers. CodeGreen achieves $R^2 = 0.9934$ correlation with mean absolute error of 10.9\%, demonstrating high measurement fidelity. The systematic offset, where CodeGreen reports slightly higher values, arises from checkpoint instrumentation introducing additional computation and CodeGreen measuring energy between explicit checkpoints while perf captures total process energy including initialization. The strong correlation confirms that CodeGreen accurately tracks energy consumption patterns without measurement validity being compromised by instrumentation overhead.

\noindent \emph{\textbf{Linearity and Scalability}}
Energy consumption should scale linearly with computational work for constant-complexity algorithms~\cite{rajput2024enhancing}. We verify this by executing benchmarks with increasing workload sizes. CodeGreen achieves $R^2 = 0.9997$ for energy-workload linearity, confirming that the asynchronous buffer management introduces no non-linear artifacts. This near-perfect linearity validates that the lock-free circular buffer architecture maintains measurement fidelity even under extreme checkpoint densities, with no buffer contention or scaling noise as workload intensity increases.

\noindent \emph{\textbf{Cross-Language Consistency}}
We validate language extensibility by comparing energy profiles across Python, C, C++, and Java. Measurements reveal expected efficiency orderings: compiled languages consume significantly less energy than interpreted languages, with Java falling in the middle due to JIT compilation. Normalized overhead remains consistent within language families after accounting for fixed initialization and per-checkpoint costs. For example, coarse-grained instrumentation shows 9.1\% overhead for C and near-zero for Python, while fine-grained instrumentation with 4 checkpoints incurs 1.5\% (C) to 11.3\% (C++) overhead. Despite native execution time differences spanning three orders of magnitude (40ms for C vs 2.92s for Python in \textbf{nbody}), CodeGreen provides consistent, predictable measurements across all tested languages, validating the architecture's language-agnostic design.

\section{Implications}
CodeGreen's design enables energy optimization workflows across multiple stakeholders. \textbf{For developers}, IDE integration would enable early-stage optimization by profiling energy during local development, highlighting hotspots through inline annotations, and preventing energy regressions before production. 
\textbf{For practitioners}, the platform eliminates tooling fragmentation, providing a consistent methodology across heterogeneous stacks (e.g., Python ML, C++ inference, GPUs) to replace disparate, incompatible profilers. 
\textbf{For researchers}, the structured JSON output facilitate LLM-driven ``measure-optimize-validate'' loops for AI-driven green software engineering. 
The deterministic overhead model ensures reproducible measurements for comparative studies, while extensibility supports emerging hardware and languages without core modifications. 
\textbf{For quality assurance teams}, fine-grained energy regression testing enables method-level budget assertions, pinpointing specific functions causing increased consumption rather than detecting only application-level regressions. By supporting commodity hardware with automated instrumentation, CodeGreen democratizes energy optimization, making sustainability a tractable engineering objective for startups, small teams, and educational contexts where energy efficiency can be taught alongside time complexity.


\section{Threats to Validity}
\textbf{Construct Validity.} Reliance on RAPL as ground truth imposes a 1ms resolution limit and platform-specific measurement characteristics. We mitigated this through long-running workloads and validated strong correlation with RAPL to confirm physical validity. Linear interpolation between samples introduces estimation error for sub-millisecond code blocks, though near-perfect energy-workload linearity suggests no systematic bias. \textbf{Internal Validity.} The extreme overhead in \textbf{nbody} with 1M checkpoints represents an intentional stress test demonstrating per-invocation measurement capability; standard coarse-grained usage incurs negligible overhead. The deterministic scaling validates that overhead is a configurable trade-off rather than stochastic measurement noise. \textbf{External Validity.} Experiments focus on x86 architecture, though CodeGreen's modular \texttt{EnergyProvider} interface supports extensibility to AMD, NVIDIA, and ARM platforms. Consistent measurements across Python, C, C++, and Java demonstrate robust applicability across diverse ecosystems, though validation on more architectures would strengthen generalizability claims.

\section{Conclusions}

CodeGreen addresses critical limitations in software energy profiling through an asynchronous producer-consumer architecture that decouples instrumentation from measurement. Achieving low overhead for whole-program profiling while enabling fine-grained method-level attribution at predictable cost. The modular design provides hardware and language extensibility supporting diverse execution environments. Validation demonstrates strong correlation with RAPL ground truth and near-perfect energy-workload linearity across multiple languages and computational patterns. By bridging coarse-grained monitoring and manual instrumentation, CodeGreen provides developers with practical infrastructure for algorithmic energy optimization across heterogeneous computing environments.

\bibliographystyle{ACM-Reference-Format}
\bibliography{ref}

%
\appendix

\section{Appendix: Tool Demonstration Walkthrough}
\label{sec:appendix_demo}

This appendix provides a step-by-step walkthrough of the CodeGreen command-line interface (CLI), demonstrating core capabilities from installation to fine-grained energy measurement. We present a practical usage guide with screenshots showing each stage of a typical workflow.

\subsection{Availability}
CodeGreen is an active, open-source project available through the following resources:
\begin{itemize}
    \item \textbf{Website \& Documentation:} \url{https://smart-dal.github.io/codegreen/}
    \item \textbf{Source Code:} \url{https://github.com/SMART-Dal/codegreen}
    \item \textbf{Video Demonstration:} \url{https://smart-dal.github.io/codegreen/demo/}
    \item \textbf{License:} Mozilla Public License 2.0 (MPL 2.0)
\end{itemize}

\begin{figure}[!htb]
    \centering
    \includegraphics[width=\columnwidth]{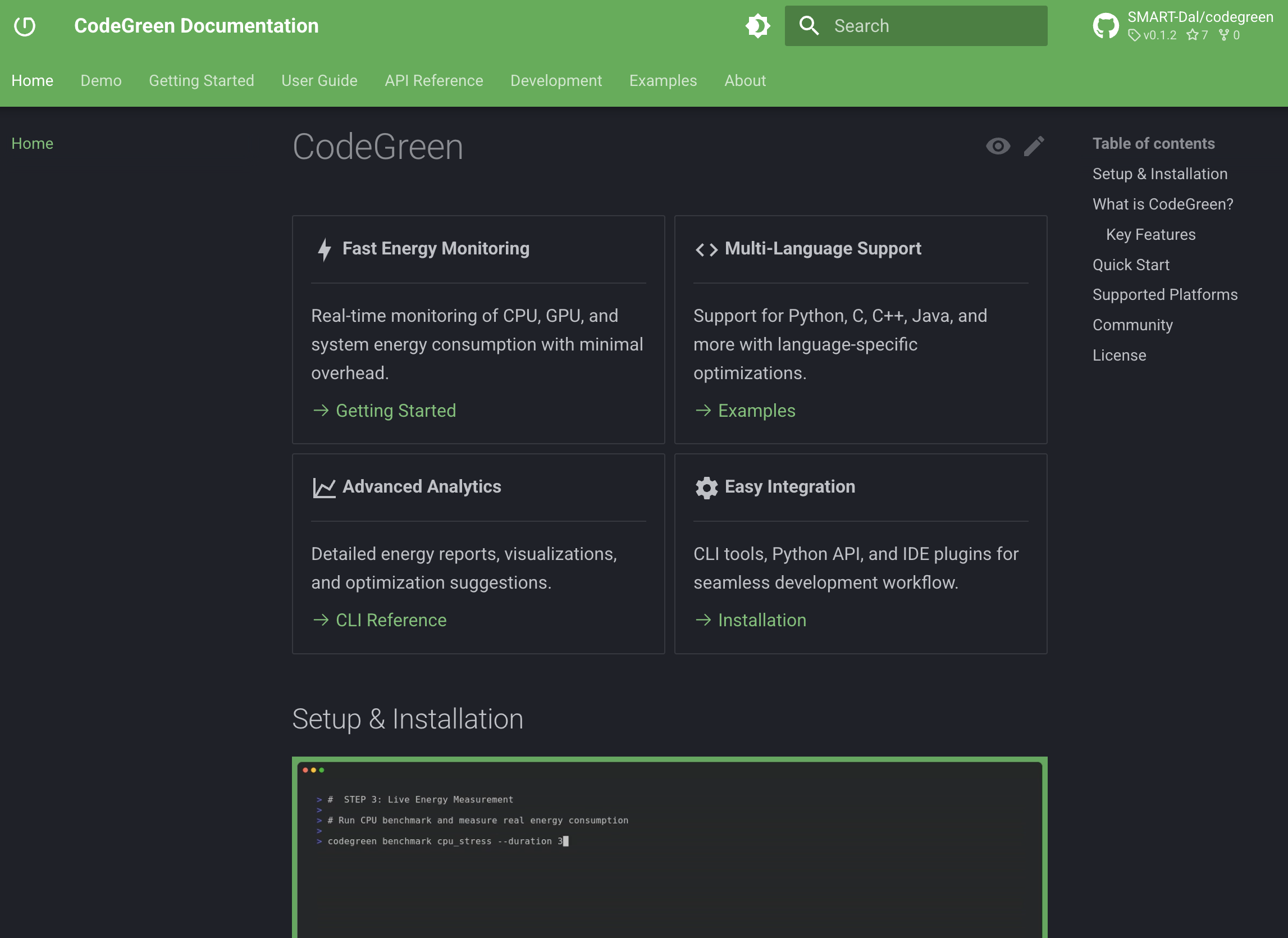}
    \caption{\textbf{CodeGreen Website:} The official website at \url{https://smart-dal.github.io/codegreen/} provides installation guides, API references, usage examples, and demo videos.}
    \label{fig:website}
\end{figure}

The project website (Fig.~\ref{fig:website}) provides comprehensive documentation, video demonstrations, installation guides, and API references to support users from initial setup through advanced usage scenarios.

\subsection{Quick Start Guide}

\noindent Table~\ref{tab:demo_map} provides a navigation index for the demonstration figures.

\begin{table}[htb]
    \centering
    \caption{Navigation Index for CodeGreen Demonstration}
    \vspace{-2mm}
    \label{tab:demo_map}
    \small
    \renewcommand{\arraystretch}{1}
    \begin{tabular}{@{} l l p{0.45\columnwidth} @{}} 
        \toprule
        \textbf{Fig.} & \textbf{Stage} & \textbf{Purpose} \\
        \midrule
        \ref{fig:install} & Install & Setup \& Compilation \\
        \ref{fig:help} & Help & Explore subcommands \\
        \ref{fig:init} & Init & Detect hardware sensors \\
        \ref{fig:info} & Info & Verify system config \\
        \ref{fig:benchmark} & Bench & Live stress testing \\
        \ref{fig:analyze} & Analyze & Static AST instrumentation \\
        \ref{fig:measure} & Measure & Fine-grained profiling \\
        \ref{fig:json} & JSON & CI/CD data export \\
        \ref{fig:config1} & Config & Tuning granularities \\
        \ref{fig:doctor} & Doctor & System diagnostics \\
        \bottomrule
    \end{tabular}
    \vspace{-2mm}
\end{table}

\begin{figure*}[!htb]  
    \centering
    \includegraphics[width=0.8\textwidth]{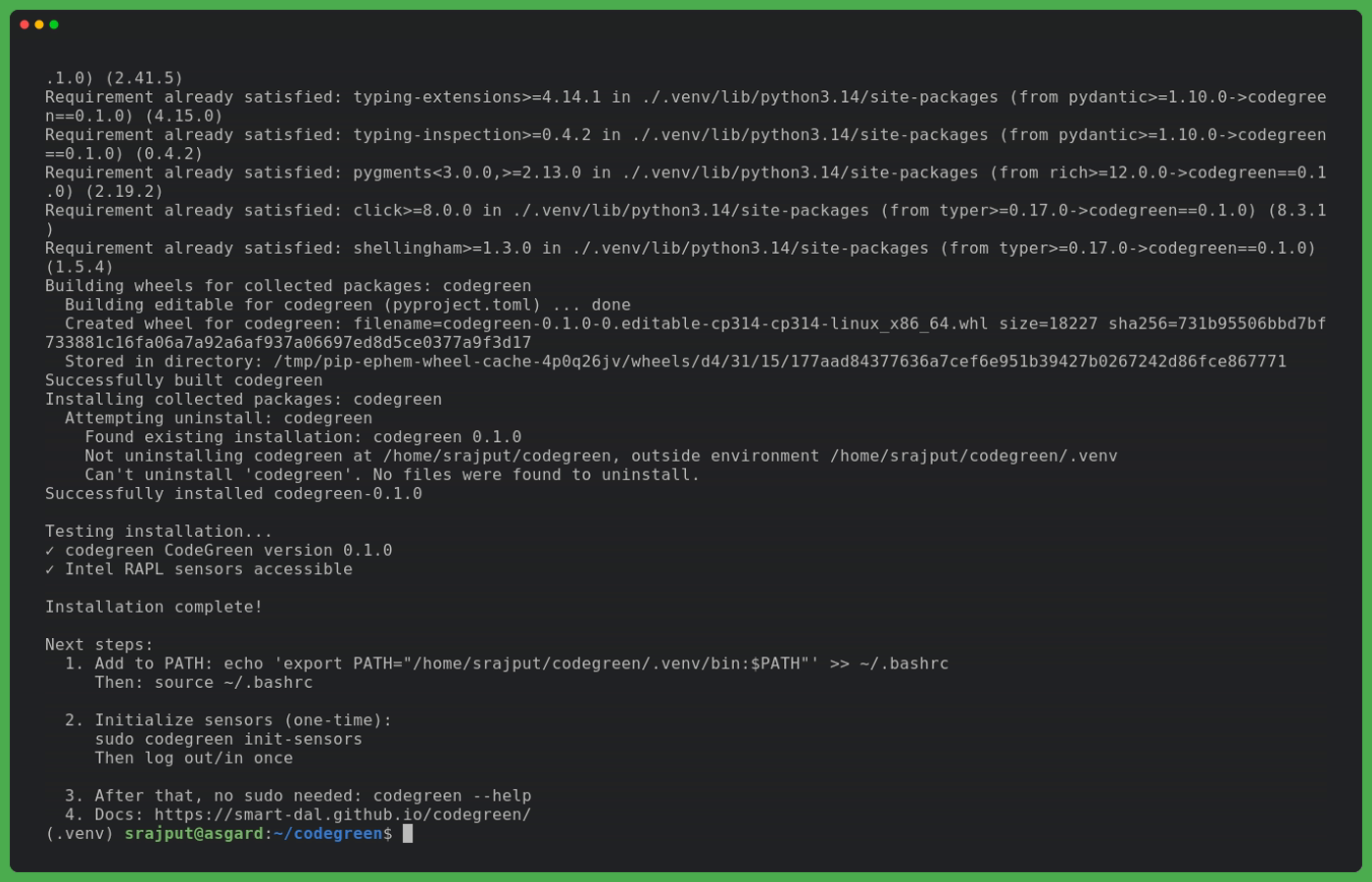}
    \caption{\textbf{Installation:} Running \texttt{./install.sh} automatically detects the host architecture, compiles NEMB, and installs Python bindings and CLI tools.}
    \label{fig:install}
\end{figure*}

\paragraph{\textbf{Step 1: Installation}}
Clone the repo and run installation script:
\begin{lstlisting}[language=bashstyle]
git clone https://github.com/SMART-Dal/codegreen.git
cd codegreen
./install.sh
\end{lstlisting}

\noindent The installer automatically detects your system architecture, compiles the Native Energy Measurement Backend (NEMB), and installs Python bindings (Fig.~\ref{fig:install}).

\begin{figure*}[!htb]
    \centering
    \includegraphics[width=0.8\textwidth]{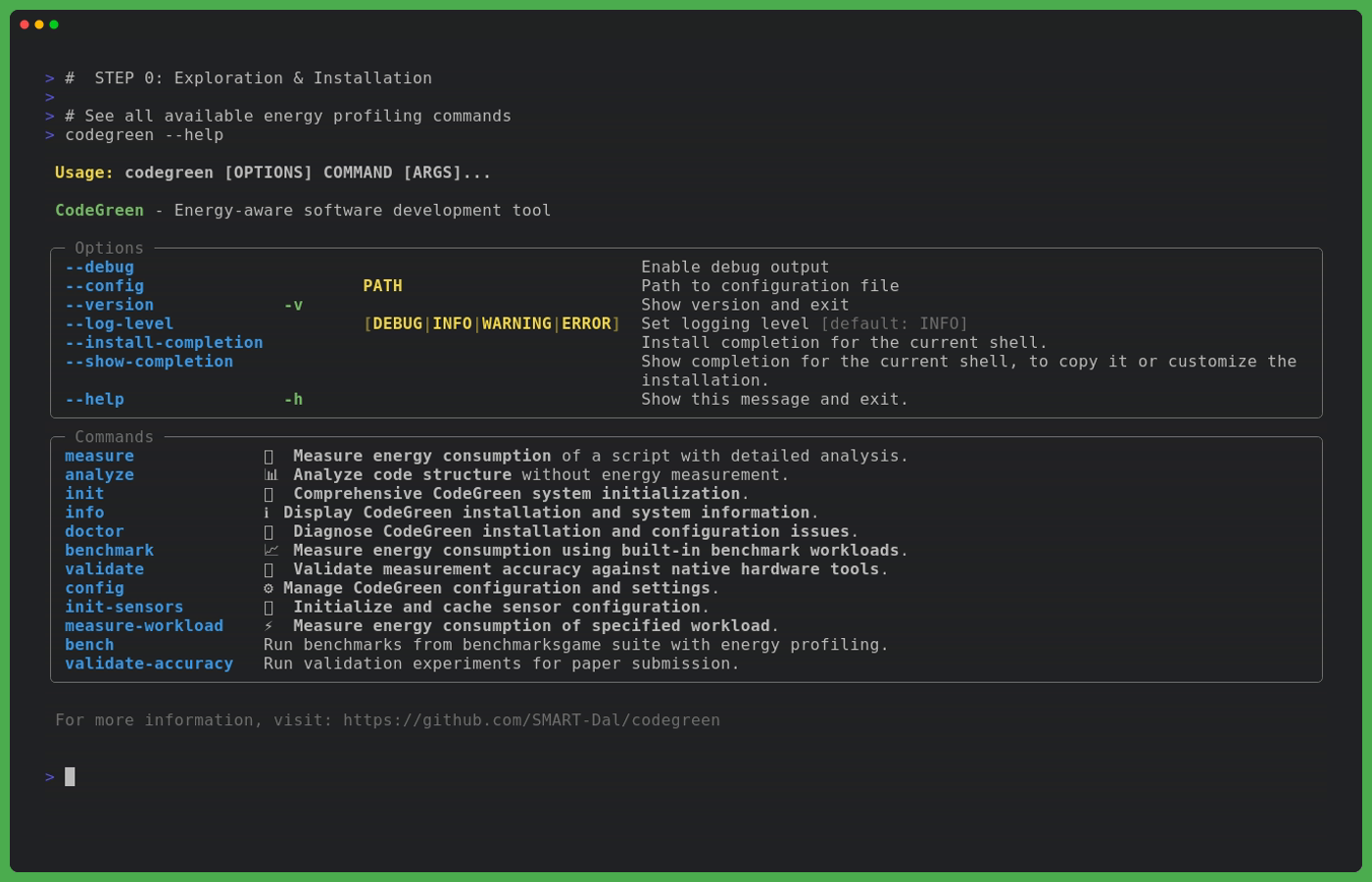}
    \caption{\textbf{Command Overview:} The \texttt{codegreen --help} command displays the complete CLI interface, categorizing subcommands into measurement (\texttt{measure}, \texttt{benchmark}), analysis (\texttt{analyze}), and system management (\texttt{init-sensors}, \texttt{config}, \texttt{doctor}, \texttt{info}) tasks.}
    \label{fig:help}
\end{figure*}

\paragraph{\textbf{Available Commands:}}After installation, explore the available CLI commands to familiarize yourself with CodeGreen's capabilities (Fig.~\ref{fig:help}):
\begin{lstlisting}[language=bashstyle]
codegreen --help
\end{lstlisting}

\noindent The help interface organizes commands into three main categories: measurement operations for profiling energy consumption, analysis tools for static code inspection, and system management utilities for configuration and diagnostics.

\begin{figure*}[!htb]
    \centering
    \includegraphics[width=0.8\textwidth]{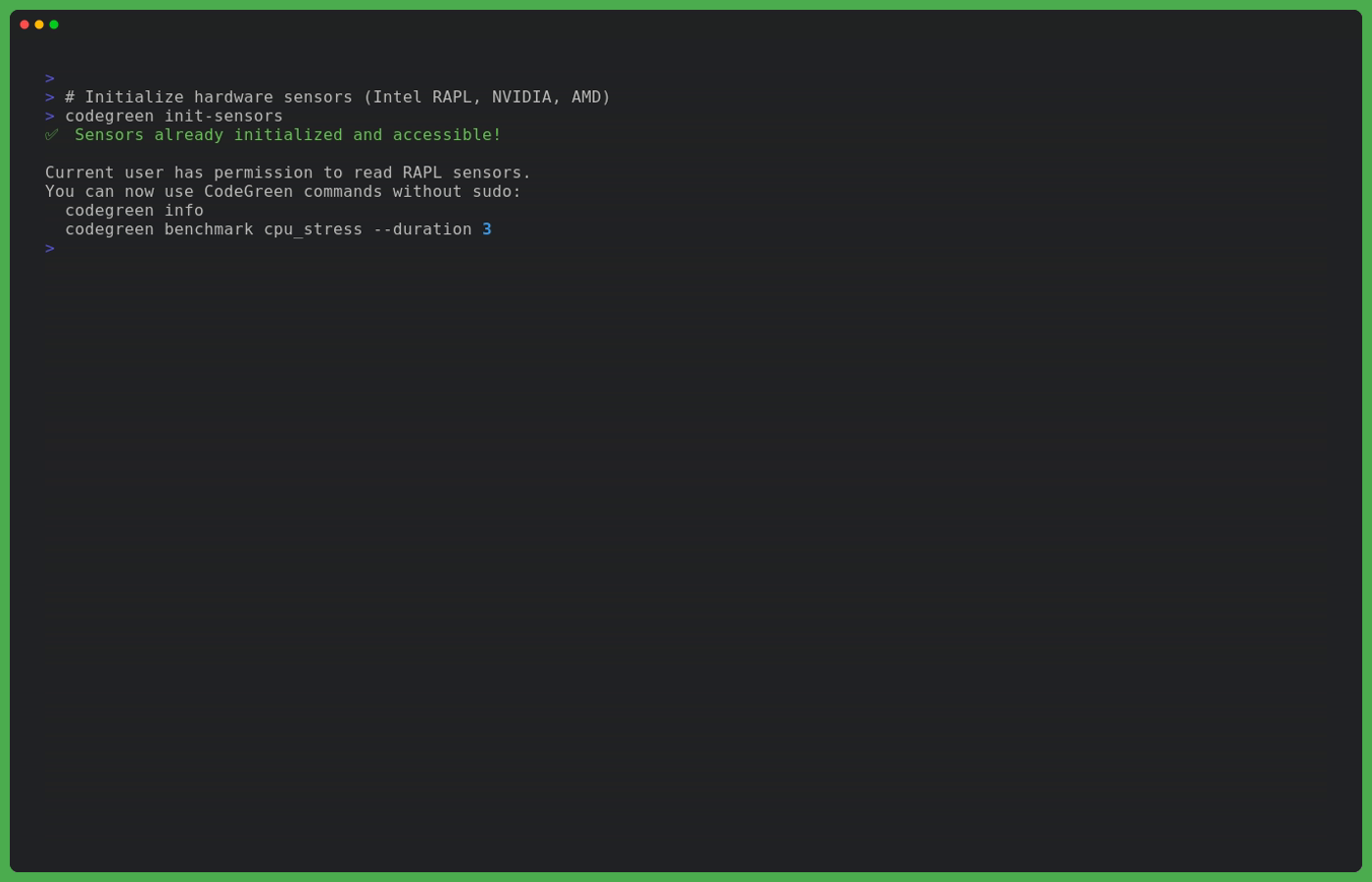}
    \caption{\textbf{Sensor Initialization:} The \texttt{codegreen init-sensors} command detects Intel RAPL, NVIDIA NVML, and AMD ROCm interfaces, configuring non-root access permissions.}
    \label{fig:init}
\end{figure*}

\paragraph{\textbf{Step 2: Initialize Hardware Sensors}}
Configure CodeGreen to detect and access available energy sensors:
\begin{lstlisting}[language=bashstyle]
codegreen init-sensors
\end{lstlisting}

\noindent This command identifies Intel RAPL, NVIDIA NVML, and AMD ROCm interfaces, configuring non-root access permissions where possible (Fig.~\ref{fig:init}).

\begin{figure*}[!htb]
    \centering
    \includegraphics[width=0.8\textwidth]{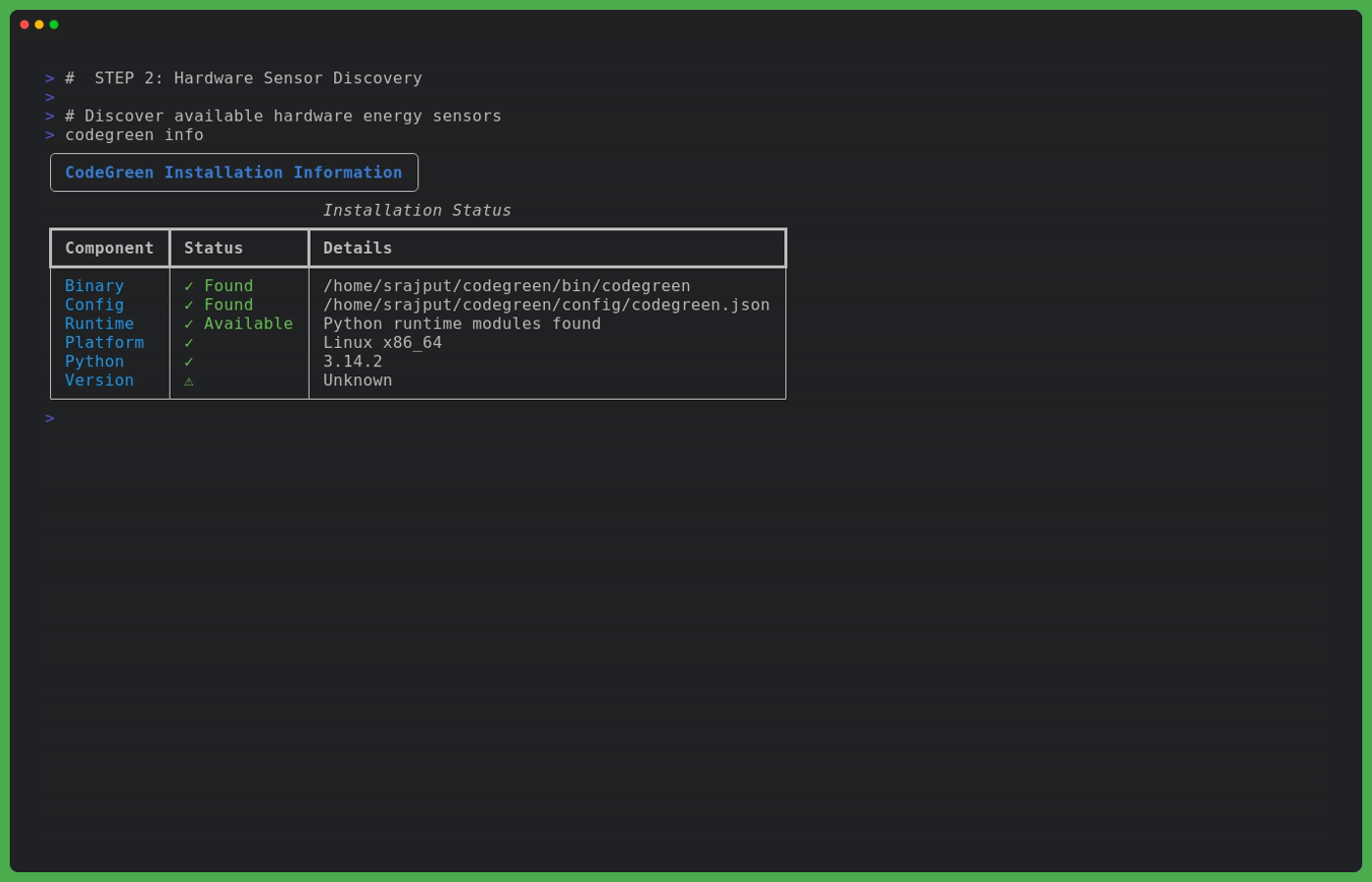}
    \caption{\textbf{System Verification:} The \texttt{codegreen info} command reports active configuration, confirming available sensors.}
    \label{fig:info}
\end{figure*}

Verify successful initialization:
\begin{lstlisting}[language=bashstyle]
codegreen info
\end{lstlisting}

\noindent The output confirms available sensors, RAPL domains (Package, Core, DRAM), and GPU devices (Fig.~\ref{fig:info}).

\begin{figure*}[!htb]
    \centering
    \includegraphics[width=0.8\textwidth]{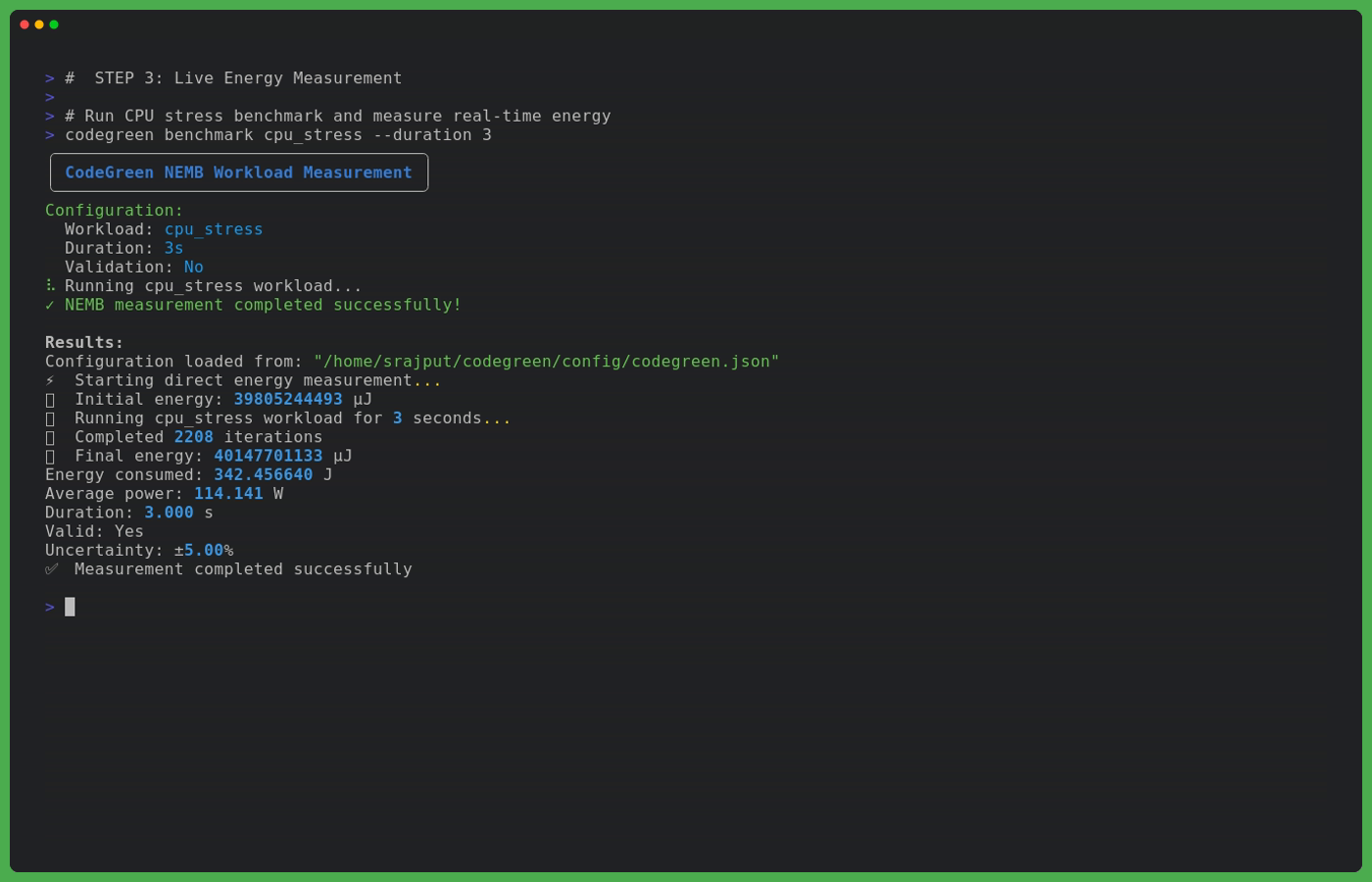}
    \caption{\textbf{Live Benchmarking:} The \texttt{codegreen benchmark} command validates the measurement pipeline with a CPU stress test.}
    \label{fig:benchmark}
\end{figure*}

\paragraph{\textbf{Step 3: Validate with Built-in Benchmark}}
Test the measurement pipeline with a CPU stress workload:
\begin{lstlisting}[language=bashstyle]
codegreen benchmark --duration 10
\end{lstlisting}

\noindent This reports total energy consumption (Joules), average power (Watts), and execution time with confidence intervals (Fig.~\ref{fig:benchmark}).

\begin{figure*}[!htb]
    \centering
    \includegraphics[width=0.8\textwidth]{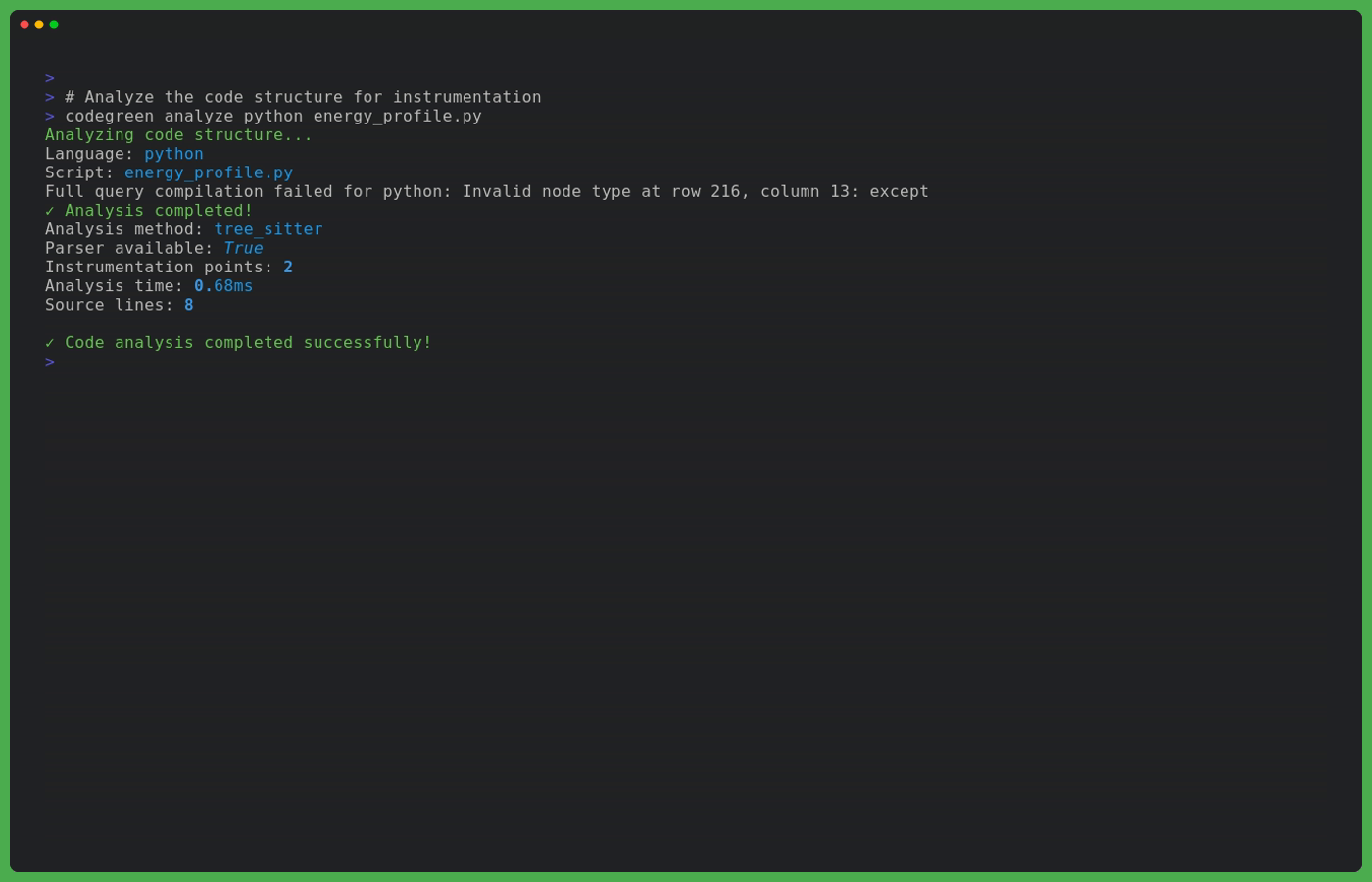}
    \caption{\textbf{Static Analysis:} The \texttt{codegreen analyze script.py} command uses Tree-sitter AST queries to identify instrumentation points (functions, loops, classes) based on user-configured granularity settings, without executing code.}
    \label{fig:analyze}
\end{figure*}

\paragraph{\textbf{Step 4: Analyze Source Code}}
Before instrumenting, preview what will be measured using static analysis:
\begin{lstlisting}[language=bashstyle]
codegreen analyze script.py
\end{lstlisting}

\noindent Based on user-specified granularity in the configuration (e.g., functions, loops, or classes), Tree-sitter AST queries~\cite{max_brunsfeld_2025_17921700} identify and target the corresponding code constructs for instrumentation without executing the program (Fig.~\ref{fig:analyze}). This automated analysis provides developers with granular control over measurement scope while eliminating manual checkpoint placement.

\begin{figure*}[!htb]
    \centering
    \includegraphics[width=0.8\textwidth]{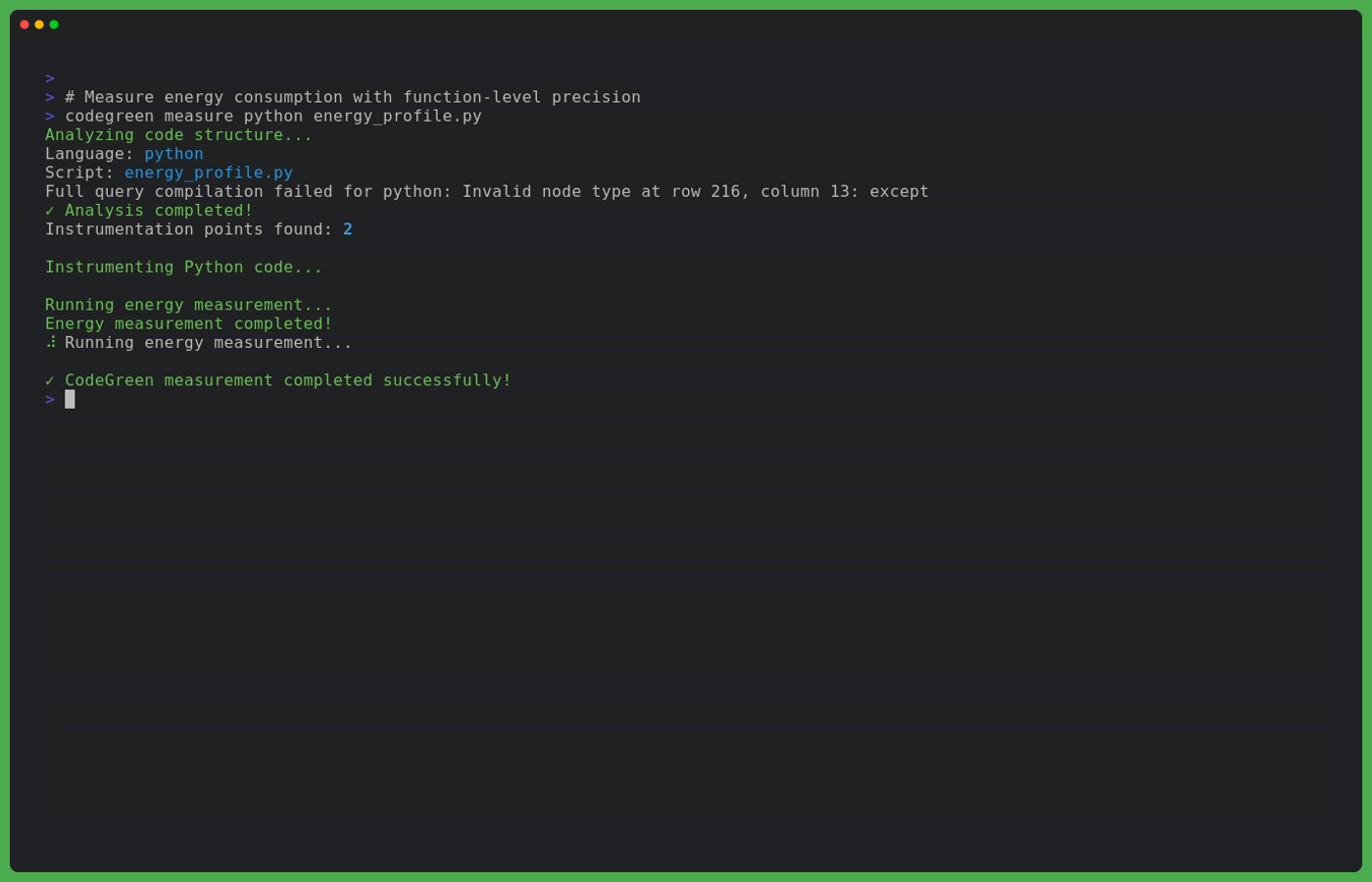}
    \caption{\textbf{Fine-Grained Measurement:} The \texttt{codegreen measure} command instruments code and attributes energy to specific functions.}
    \label{fig:measure}
\end{figure*}

\paragraph{\textbf{Step 5: Measure Energy Consumption}}
Execute fine-grained profiling with automatic instrumentation:
\begin{lstlisting}[language=bashstyle]
codegreen measure script.py
\end{lstlisting}

\noindent CodeGreen instruments the code, runs it, and attributes energy to specific functions, identifying hotspots (Fig.~\ref{fig:measure}).

\begin{figure*}[!htb]
    \centering
    \includegraphics[width=0.8\textwidth]{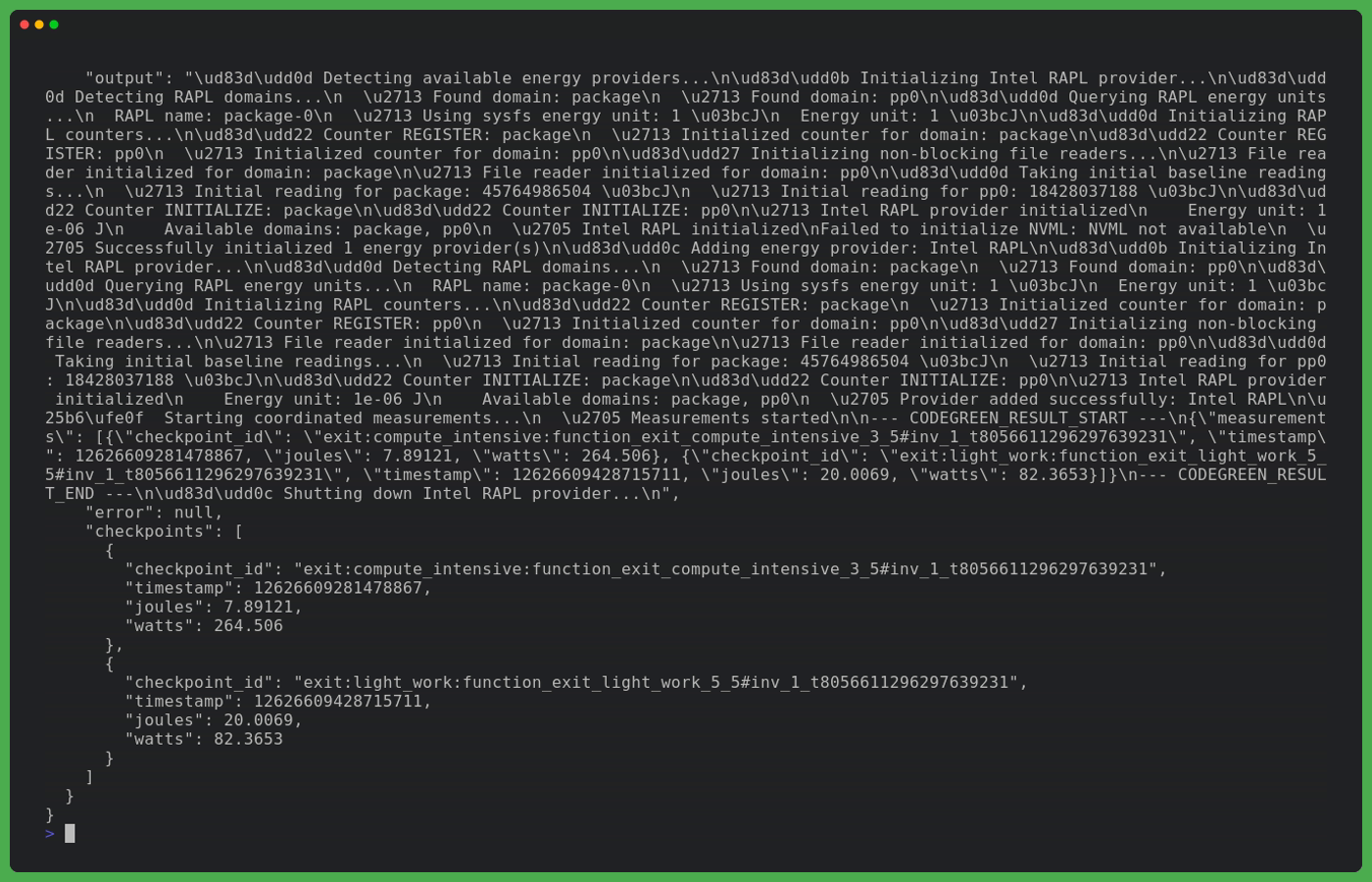}
    \caption{\textbf{JSON Export:} Using \texttt{--output json} enables CI/CD integration for automated energy regression testing.}
    \label{fig:json}
\end{figure*}

\noindent For CI/CD integration, export results as JSON:
\begin{lstlisting}[language=bashstyle]
codegreen measure script.py --output json
\end{lstlisting}

\noindent This structured format enables automated regression testing (Fig.~\ref{fig:json}).

\begin{figure*}[!htb]
    \centering
    \includegraphics[width=0.8\textwidth]{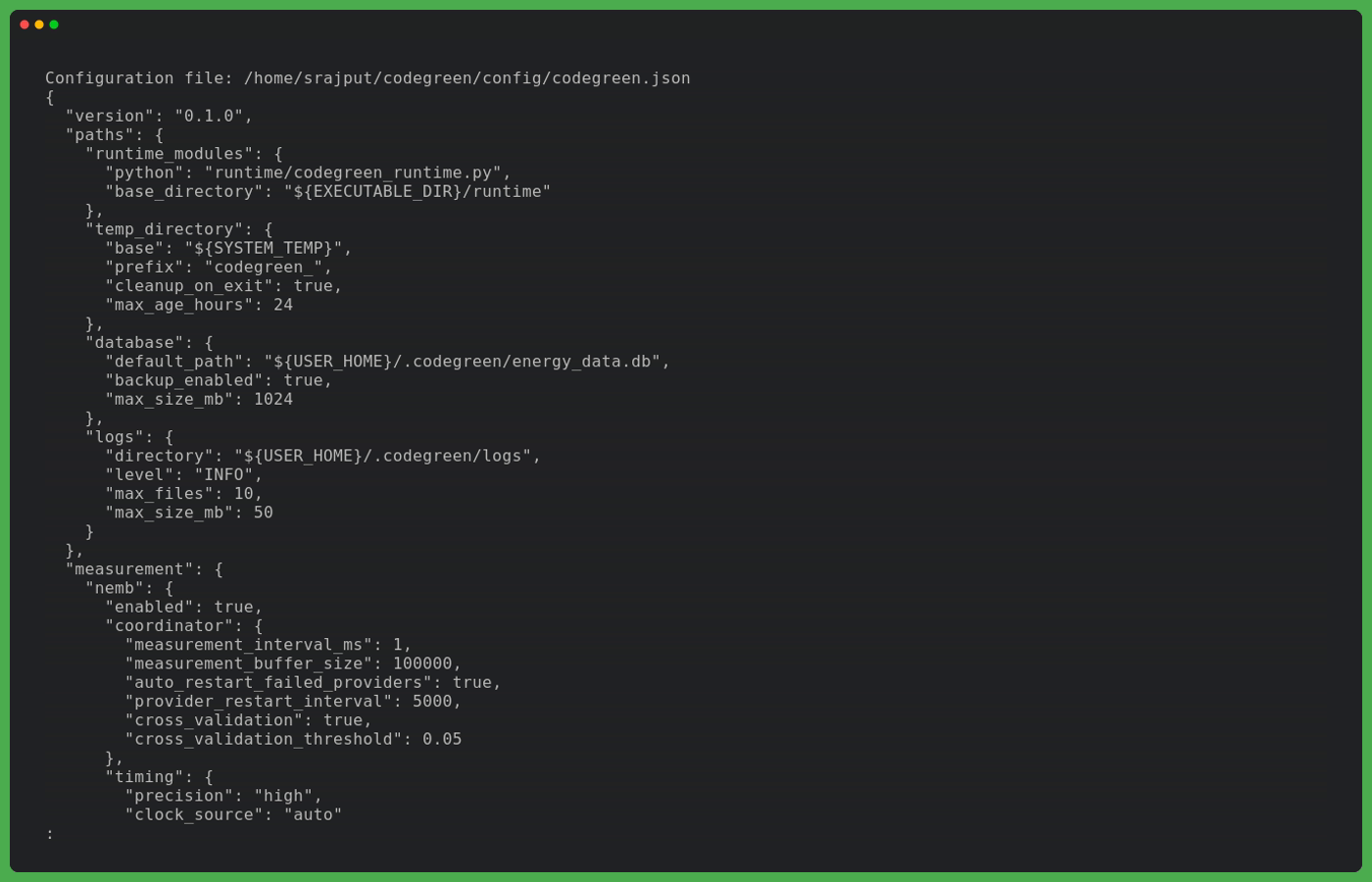}
    \caption{\textbf{Configuration:} The \texttt{codegreen config} command tunes sensor polling intervals and instrumentation granularity.}
    \label{fig:config1}
\end{figure*}

\paragraph{\textbf{Step 6: Configure Measurement Granularity}}
Tune profiling behavior to balance precision and overhead:
\begin{lstlisting}[language=bashstyle]
codegreen config
\end{lstlisting}

Adjust sensor polling intervals, accuracy thresholds, and instrumentation granularity (Fig.~\ref{fig:config1}).

\begin{figure*}[!htb]
    \centering
    \includegraphics[width=0.8\textwidth]{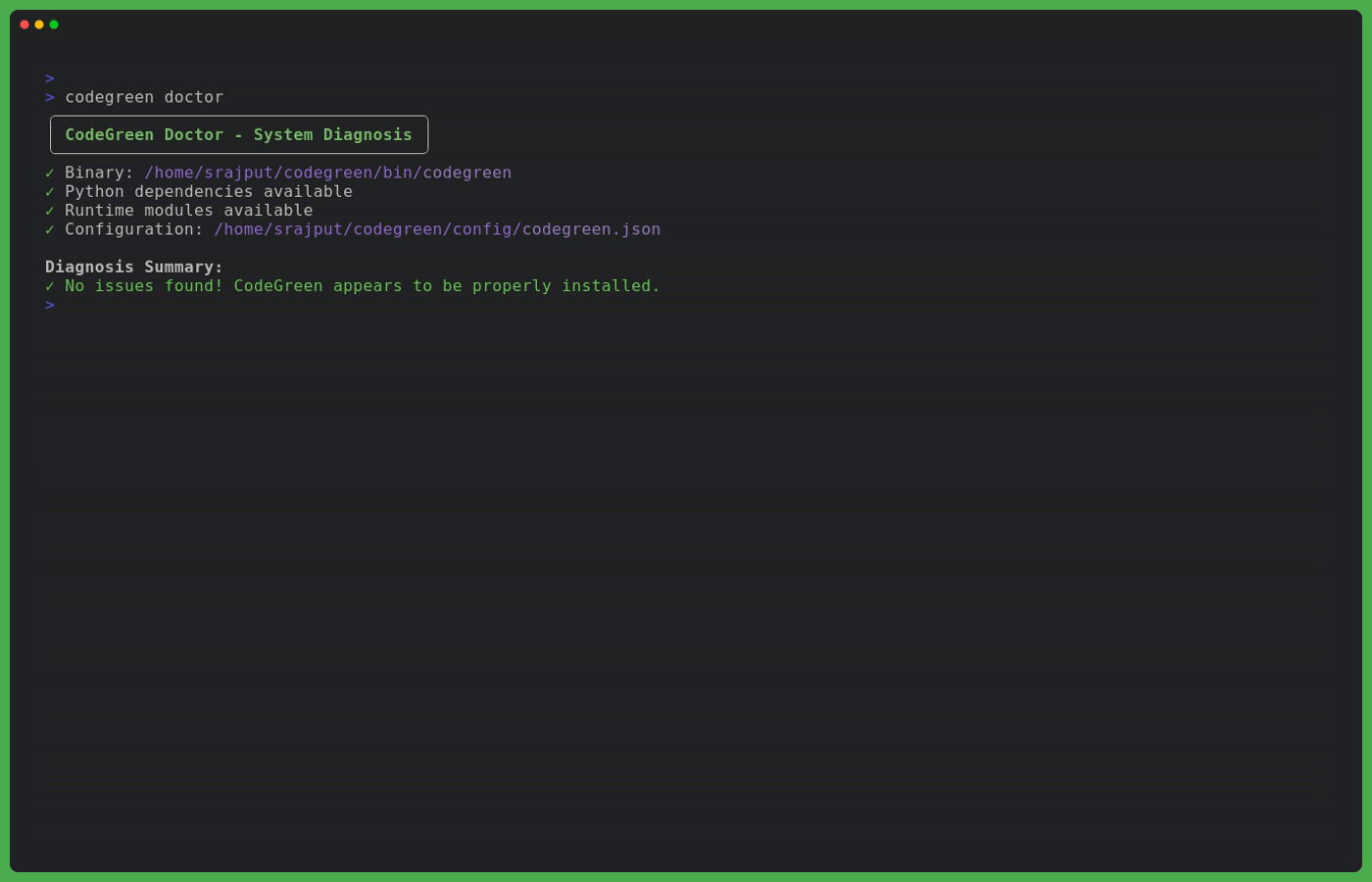}
    \caption{\textbf{Diagnostics:} The \texttt{codegreen doctor} command verifies file permissions, validates dependencies, and tests sensor health.}
    \label{fig:doctor}
\end{figure*}

\paragraph{\textbf{Step 7: System Diagnostics}}
Verify installation integrity and sensor health:
\begin{lstlisting}[language=bashstyle]
codegreen doctor
\end{lstlisting}

This checks file permissions, validates dependencies, and tests sensor read operations (Fig.~\ref{fig:doctor}).

\subsection{Typical Workflow Summary}
A typical energy optimization workflow is as follows:
\begin{enumerate}
    \item \textbf{Initialize:} Detect hardware sensors with \texttt{init-sensors} and verify configuration using \texttt{info}
    \item \textbf{Baseline:} Establish baseline energy consumption using \texttt{benchmark} to validate measurement pipeline
    \item \textbf{Profile:} Identify energy hotspots by running \texttt{measure script} with appropriate granularity settings
    \item \textbf{Optimize:} Refactor hotspots based on profiling results
    \item \textbf{Validate:} Re-measure optimized code to quantify energy improvements and prevent regressions
    \item \textbf{Automate:} Integrate JSON output into CI/CD pipelines for continuous energy regression testing
\end{enumerate}

\end{document}